\documentclass[twocolumn,showpacs,prb]{revtex4}% Physical Review B
\usepackage{graphicx}% Include figure files
\usepackage{dcolumn}% Align table columns on decimal point
\usepackage{bm}% bold math
\usepackage{epsfig}
\begin{document}
%\preprint{APS/123-QED}
\title{Neutron scattering study of the vibrations in vitreous silica and germania}% Force line breaks with \\
\author{E. Fabiani}
\affiliation{%
Institute Laue-Langevin, BP 156, F-38042 Grenoble Cedex 9, France
}%
\author{A. Fontana}
\affiliation{%
Dipartimento di Fisica, Universit$\acute{a}$ di Trento, I-38050,
Povo (Trento), Italy and INFM CRS-SOFT, c/o Universit$\acute{a}$
di Roma "La Sapienza", I-00185, Roma, Italy
}%
\author{U. Buchenau}
 \email{buchenau-juelich@t-online.de}
\affiliation{%
Institut f\"ur Festk\"orperforschung, Forschungszentrum J\"ulich\\
Postfach 1913, D--52425 J\"ulich, Federal Republic of Germany
}%
\date{February 8, 2005}% It is always \today, today,
\begin{abstract}
The incoherent approximation for the determination of the
vibrational density of states of glasses from inelastic neutron or
x-ray scattering data is extended to treat the coherent
scattering. The method is applied to new room temperature
measurements of vitreous silica and germania on the thermal
time-of-flight spectrometer IN4 at the High Flux Reactor in
Grenoble. The inelastic dynamic structure factor at the boson peak
turns out to be reasonably well described in terms of a mixture of
rotation and translation of practically undistorted SiO$_4$ or
GeO$_4$ tetrahedra. The translational component exceeds the
expectation of the Debye model by a factor of two. A possible
relation of this excess to the phonon shift and broadening
observed in x-ray Brillouin scattering experiments is discussed.
\end{abstract}
\maketitle
\section{Introduction}
There is an extensive quantum mechanical treatment of the
scattering from atoms moving in a crystal \cite{lovesey}. The
regular atomic arrangement allows to solve this problem with an
accuracy and a theoretical depth which one cannot hope for in the
much more complex case of a disordered solid. In a glass, the
translation symmetry is lost. In addition, the sample is in an
energy landscape with many minima. Fortunately, the relaxational
jumps between different energy minima of glasses are only seen as
quasielastic scattering below an energy transfer of about 1 meV
\cite{sokolov,wiedersich} (a frequency of 250 GHz); above that
frequency, one can reckon with a more or less harmonic vibrational
density of states.

But the problem what these vibrations are is by no means solved.
Below 1 meV, reasonably well-defined long-wavelength sound waves
can be shown to exist (coexisting with relaxational or tunneling
motion), but they become rapidly overdamped above that frequency.
This was first deduced \cite{zepo} from the plateau in the thermal
conductivity at low temperatures. Within the last decade, it has
been directly observed for the longitudinal sound waves by x-ray
Brillouin scattering \cite{sette}. Also, the density of states
exceeds the Debye expectation markedly at the so-called boson
peak, at a frequency where the corresponding crystals still have
only well-defined long-wavelength sound waves.

In this paper, we report thermal neutron time-of-flight
measurements at room temperature on silica and germania, data
taken over a very large momentum transfer range with the new
spectrometer IN4 at the High Flux Reactor at the Institut
Laue-Langevin at Grenoble. The high quality of the data allows for
an evaluation which goes beyond the usual incoherent
approximation, extending and improving earlier work
\cite{bu,price,wischi,arai1,arai2}. One not only gets a
vibrational density of states which compares favorably with heat
capacity data, but one learns new facts on the details of the
atomic motion, in particular in the low-frequency range at the
boson peak.

After this introduction, the paper describes an extension of the
incoherent approximation for the evaluation of coherent inelastic
neutron or x-ray data in Section II. Section III presents the
time-of-flight experiments in silica and germania, their
multiple-scattering correction and their normalization to
diffraction data from the literature. Section IV applies the new
method to the data. The discussion of the results and a short
summary is given in Sections V and VI.

\section{Extending the incoherent approximation}

\subsection{Definitions}

The following derivation is formulated in terms of the classical
scattering law $S(Q,\omega)$, where the frequency $\omega$ is
related to the energy transfer $E$ of the scattering process by
$E=\hbar\omega$ and $Q$ is the momentum transfer. This classical
$S(Q,\omega)$ can be calculated from the measured double
differential cross section
\begin{equation}\label{sqom}
S(Q,\omega)=\frac{k_BT}{\hbar\omega}({\rm
e}^{\hbar\omega/k_BT}-1)\frac{k_i}{k_f}\frac{4\pi}{N\overline{\sigma}}
\frac{d^2\sigma}{d\omega d\Omega},
\end{equation}
taking $\omega$ to be positive in energy gain of the scattered
particle, neutron or x-ray photon. Here $T$ is the temperature,
$k_i$ and $k_f$ are the wavevector values of incoming and
scattered waves, respectively, $N$ is the number of atoms in the
beam, $\overline{\sigma}$ is the average scattering cross section
of the atoms and $\Omega$ is the solid angle. Note that the
average scattering cross section $\overline{\sigma}$ is
$Q$-dependent for x-rays; this $Q$-dependence is determined by the
atomic form factors. The definition of $S(Q,\omega)$ requires a
completely isotropic glass, the usual case. It is also valid for
glasses with more than one kind of atom, like silica and germania.

The above definition makes no distinction between coherent
scattering (the case where the waves scattered from different
atoms interfere) and incoherent scattering (the case where there
is no interference). X-ray scattering is purely coherent; for
neutrons, it depends on the nuclei of the scattering atoms. In
silicon, germanium and oxygen atoms, the coherent scattering
dominates. Integrating over all frequencies one obtains $S(Q)$
\begin{equation}
S(Q)\equiv\int_{-\infty}^\infty d\omega S(Q,\omega).
\end{equation}
If all atoms of the sample scatter only incoherently, $S(Q)=1$. If
one has only coherent scattering, $S(Q)$ shows a first sharp
diffraction peak at about 1.5 ${\rm \AA}$, followed by
oscillations around 1 which die out at high $Q$. These peaks
reflect the short range order of the glass on the atomic scale.
Below the first sharp diffraction peak, S(Q) is due to long-range
density and concentration fluctuations. Silica and germania show a
pronounced first sharp diffraction peak.

For the evaluation of coherent scatterers, we will need the
definition of a hypothetical incoherent scattering function
$S_{inc}(Q,\omega)$. This is defined as the scattering function
which one would have without interference between different atoms,
keeping their total cross sections.

\subsection{The incoherent approximation}

To determine a vibrational density of states from scattering data,
one needs to take the Debye-Waller factor and the multiphonon
scattering into account. An elegant way to do this is to use the
intermediate scattering function
\begin{equation}\label{fourier}
S(Q,t)=\int_{-\infty}^\infty\cos\omega t\ S(Q,\omega) d\omega.
\end{equation}
with the back transform
\begin{equation} S(Q,\omega)=
\frac{1}{\pi}\int_{0}^\infty\cos\omega t\ S(Q,t) dt,
\end{equation}

The incoherent approximation assumes that one can describe the
scattering function, eq. (\ref{sqom}), in terms of an average atom
which scatters only incoherently. The time-dependent displacement
of this average atom from its equilibrium position is assumed to
have a Gaussian distribution. From the Bloch identity
\cite{lovesey}, one obtains the intermediate scattering function
\begin{equation}\label{gauss}
S_{inc}(Q,t)= {\rm e}^{-Q^2\gamma(t)},
\end{equation}
where $\gamma(t)$ is the time-dependent mean square displacement
of the average atom.

If there are only vibrations, $\gamma(t)$ is determined by the
vibrational density of states $g(\omega)$. Their relation can be
derived from the one-phonon approximation for the inelastic
scattering from our classical isotropic incoherent scatterer
\cite{lovesey}
\begin{equation}\label{onepho}
S_{inc}(Q,\omega)= Q^2{\rm e}^{-2W}
\frac{k_BT}{2\overline{M}}\frac{g(\omega)}{\omega^2}
\end{equation}
where ${\rm e}^{-2W}$ is the Debye-Waller factor and
$\overline{M}$ is the average atomic mass. Comparing the initial
$Q^2$ rise and using the Fourier transformation, eq.
(\ref{fourier}), one finds
\begin{equation}\label{gamt}
\gamma(t)=\frac{k_BT}{\overline{M}} \int_0^{\omega_{max}}d\omega
\frac{g(\omega)}{\omega^2} (1-\cos\omega t).
\end{equation}
Here $\omega_{max}$ is the upper frequency boundary of the
vibrational density of states.

The incoherent scattering is obtained from the Fourier transform
of the  intermediate scattering function in time
\begin{equation}\label{inc}
S_{inc}(Q,\omega)= \frac{1}{\pi} \int_0^\infty dt\cos\omega t{\rm
e}^{-\gamma(t)Q^2}.
\end{equation}
In this approximation, one accounts not only for the one-phonon
scattering, but the entire inelastic scattering including the
multiphonon terms.

To apply the approximation to measured data, one begins by
calculating a first guess to the vibrational density of states,
assuming a $Q^2$- or $Q^2{\rm e}^{-2W}$-dependence of the
inelastic scattering. From the density of states, one gets
$\gamma(t)$ from eq. (\ref{gamt}) and can calculate the
$Q$-dependence of the inelastic scattering. With this, one can
determine a better density of states from the data. Usually, after
a few iterations the density of states does no longer change. In
the case of a glass consisting of two or more elements, one calls
this density the {\it generalized vibrational density of states},
to emphasize that is is not the true density of states, but its
reflection in the scattering, weighted by the cross sections and
masses of the atoms in the sample.

As we will see, the incoherent approximation works astonishingly
well even for the two coherently scattering glasses silica and
germania. But it does not provide any information on the
vibrational eigenvectors. One needs an extension of the incoherent
approximation to do that. Such an extension will be introduced in
the following subsection.

\subsection{Oscillation function $S_\omega(Q)$}

The vibrational density of states is a function of the frequency,
not of the time. Therefore the extension of the incoherent
approximation must be done in the frequency domain. The
vibrational eigenvectors change with changing frequency, so each
frequency window has its own coherent dynamic structure factor.
The interference between different scattering atoms does not
change the overall scattering intensity, but leads to oscillations
in the dependence on the momentum transfer.

To take this into account, we define the oscillation function
$S_\omega(Q)$ by
\begin{equation}\label{somq}
S_\omega(Q)=\frac{S(Q,\omega)}{S_{inc}(Q,\omega)},
\end{equation}
where $S_{inc}(Q,\omega)$ is understood to be the scattering
function without interference between different atoms, but with
unchanged total cross sections, as defined in II. A.

Like $S(Q)$, $S_\omega(Q)$ equals 1 in the incoherent case. It is
an extension of $S(Q)$ to describe the coherent oscillations of
the scattering at a fixed frequency $\omega$, both for elastic and
inelastic scattering. Though our main interest is in the inelastic
part, it is suitable to begin the discussion with the elastic part
$S_0(Q)$ for $\omega=0$.

As $S(Q)$ reflects the pair correlation of the atoms in their
instantaneous positions, the elastic oscillation function $S_0(Q)$
reflects the pair correlation of the atoms in their equilibrium
positions. In a glass at low temperatures (and even at not so low
temperatures), the atomic displacements are small compared to the
interatomic distances, so $S_0(Q)\approx S(Q)$. Both functions
show the first sharp diffraction peak at about 1.5 ${\rm\AA}$
followed by further peaks at higher momentum transfer. At still
higher momentum transfer, the peak amplitudes diminish and the
functions $S(Q)$ and $S_0(Q)$ approach the final value of 1. Below
the first sharp diffraction peak, at small momentum transfer,
$S_0(Q)$ mirrors the frozen density and concentration fluctuations
of the glass, while $S(Q)$ mirrors both static and dynamic ones.

The inelastic part of the oscillation function $S_\omega(Q)$ with
$\omega\neq 0$ will depend on the vibrational modes at the given
frequency. At small momentum transfer, it shows the Brillouin
peak, the scattering from longitudinal sound waves. The
interference pattern at higher momentum transfer is not only due
to the positional phase factors which determine $S(Q)$, but also
to the scalar product of momentum transfer and atomic displacement
(see eq. (\ref{somq1})). Thus the inelastic part of $S_\omega(Q)$
contains information on the eigenvectors of the vibrational modes
at the frequency $\omega$.

With the help of the oscillation function, the incoherent
approximation, eq. (\ref{inc}), transforms into the extended
approximation
\begin{equation}\label{extend}
S(Q,\omega)= S_\omega(Q)\frac{1}{\pi} \int_0^\infty dt\cos\omega
t{\rm e}^{-\gamma(t)Q^2}.
\end{equation}

%%%%%%%%%%%%%%%%%%%%% begin figure %%%%%%%%%%%%%%%%%%%%%%%%%%%%%%%%%%%%%
\begin{figure}[b]
\hspace{-0cm} \vspace{0cm} \epsfig{file=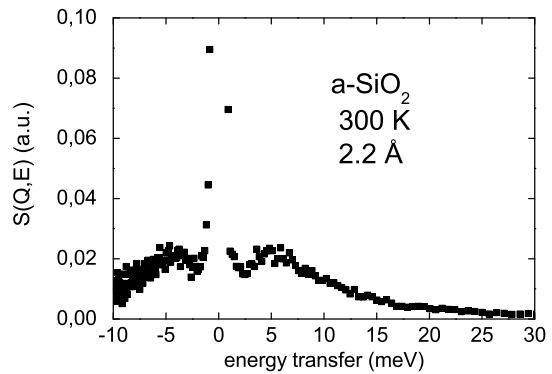,width=8
cm,angle=0} \vspace{0cm} \caption{IN4 spectrum (with standard
corrections) from vitreous silica at a scattering angle of 104
degrees, showing the boson peak at both sides of the elastic
line.}
\end{figure}
%%%%%%%%%%%%%%%%%%%%% end figure %%%%%%%%%%%%%%%%%%%%%%%%%%%%%%%%%%%%%%%

The approximation allows to fit not only a density of states, but
a frequency-dependent oscillation function as well. Thus, for
coherent scatterers, the introduction of $S_\omega(Q)$ provides
not only an enormous reduction of the deviation between incoherent
approximation and experiment, but opens up the possibility to
analyze the vibrational eigenvectors \cite{bu,wischi}. If this
analysis is successful, one can proceed to calculate the true
vibrational density of states, because then one knows the weight
of a given mode in the scattering.

In the harmonic one-phonon approximation, one can express the
coherent inelastic scattering \cite{lovesey} from a given normal
mode in terms of its eigenvector ${\bf e}_j$, the equilibrium
position ${\bf r}_j$ and the scattering length $b_j$ of atom $j$,
($j=1..N$) (the scattering length $b_j$ is related to the coherent
cross section $\sigma_j$ by $\sigma_j=4\pi b_j^2$). Within the
one-phonon approximation, the oscillation function
\begin{equation}\label{somq1}
S_\omega(Q)=\langle\frac{3}{Q^2F_{norm}}\mid\sum_{j=1}^N b_j{\rm
e}^{-i{\bf Qr}_j} \frac{{\bf Q}{\bf
e}_j}{M_j^{1/2}}\mid^2\rangle_\omega,
\end{equation}
where the angular brackets denote an average over all eigenmodes
at the frequency $\omega$, together with a directional average
over the momentum transfer vector ${\bf Q}$. The mode
normalization factor $F_{norm}$ is given by
\begin{equation} F_{norm}=
\sum_{j=1}^N \frac{b_j^2{\bf e}_j^2}{M_j}.
\end{equation}

If one is able to find the proper mode eigenvectors, one can go
beyond the determination of a generalized vibrational density of
states, because the proper normalization factor contains cross
sections and masses and thus allows to determine the true
vibrational density of states. This paper aims at such a treatment
for the silica and germania measurements described in the next
section.

\section{Experiment}

\subsection{Samples and time-of-flight experiment}

The vitreous silica sample used for the IN4 experiments was a
commercial grade spectrosil disk with a diameter of 50 mm and a
thickness of 4.8 mm.

In the case of vitreous germania, appropriate amounts of
reagent-grade GeO$_2$ powder (Aldrich 99.99+\%) were melted in
platinum crucibles for about 1 h at ca. 1600 °C. The homogeneous
and bubble-free melt was subsequently quenched in water and glassy
samples of irregular shape were removed from the bottom of the
crucible. This preparation technique results in completely
transparent glasses. Several such pieces were arranged to mimic
the disk shape of the silica sample.

The neutron spectra were obtained on the thermal time-of-flight
spectrometer IN4 at the High Flux Reactor of the Institute
Laue-Langevin at Grenoble. The measurements were done in
reflection geometry, with the sample disk plane inclined at 45
degrees to the incoming beam. With a wavelength of incoming
neutrons of 1.53 ${\rm\AA}$, one is able to study the momentum
transfer range from 1.5 to 7 ${\rm\AA}^{-1}$ with an energy
resolution of 1.3 meV FWHM. Similarly, a wavelength of incoming
neutrons of 2.2 ${\rm\AA}$ allowed to study the momentum transfer
range from 1 to 4.9 ${\rm\AA}^{-1}$ with an energy resolution of
0.8 meV FWHM.

The momentum transfer range of these measurements is more than a
factor of two larger than the one of earlier investigations
\cite{bu,wischi} with cold neutrons on the time-of-flight
spectrometer IN6. One of these earlier measurements (vitreous
silica at 318 K) was included in the evaluation presented below.

The IN4 measurements were performed at temperatures between 5 and
300 K. The following evaluation, however, is restricted to the 300
K data.

\subsection{Corrections}

The measured neutron counts were corrected for the empty container
signal. The detector efficiency was corrected for by a measurement
of a vanadium sample (an incoherent scatterer) in the same
container. The signal was multiplied by $k_i/k_f$ and an
angle-dependent absorption correction was calculated. These four
standard corrections were done with the program INX of the
Institut Laue-Langevin at Grenoble. Fig. 1 shows such a corrected
spectrum for silica at the highest scattering angle.

%%%%%%%%%%%%%%%%%%%%% begin figure %%%%%%%%%%%%%%%%%%%%%%%%%%%%%%%%%%%%%
\begin{figure}[b]
\hspace{-0cm} \vspace{0cm} \epsfig{file=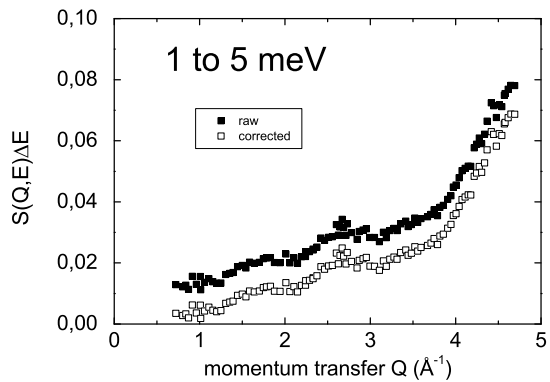,width=8
cm,angle=0} \vspace{0cm} \caption{Inelastic signal from 1 to 5 meV
for germania at 300 K, incoming wavelength 2.2 ${\rm\AA}$, with
and without multiple scattering correction.}
\end{figure}
%%%%%%%%%%%%%%%%%%%%% end figure %%%%%%%%%%%%%%%%%%%%%%%%%%%%%%%%%%%%%%%

The multiple scattering correction was done assuming an
angle-independent multiple scattering contribution. For silica and
germania, where wide-angle scattering predominates, this
assumption is expected to hold.

To do the correction, one first calculates an average spectrum,
weighting each detector with the sine of its scattering angle.
This spectrum should be folded with itself to get the spectrum of
the twice-scattering processes. However, to treat the elastic line
correctly, one has to replace the elastic line by a
$\delta$-function in one of the two spectra which one folds.
Otherwise the procedure would broaden the elastic line in an
unphysical way.

The question how much of the resulting twice-scattering spectrum
one should subtract is answered by looking at the momentum
transfer dependence in the inelastic part of the spectrum. On a
normalized scale, one subtracts a fraction $s_{mu}$, chosen in
such a way that the inelastic intensity extrapolates to zero
towards zero momentum transfer. As an example, Fig. 2 shows
corrected (with $s_{mu}=0.1$) and uncorrected data for germania at
300 K between 1 and 5 meV, measured with incoming neutrons of a
wavelength of 2.2 ${\rm\AA}$.

%%%%%%%%%%%%%%%%%%%%% begin figure %%%%%%%%%%%%%%%%%%%%%%%%%%%%%%%%%%%%%
\begin{figure}[b]
\hspace{-0cm} \vspace{0cm} \epsfig{file=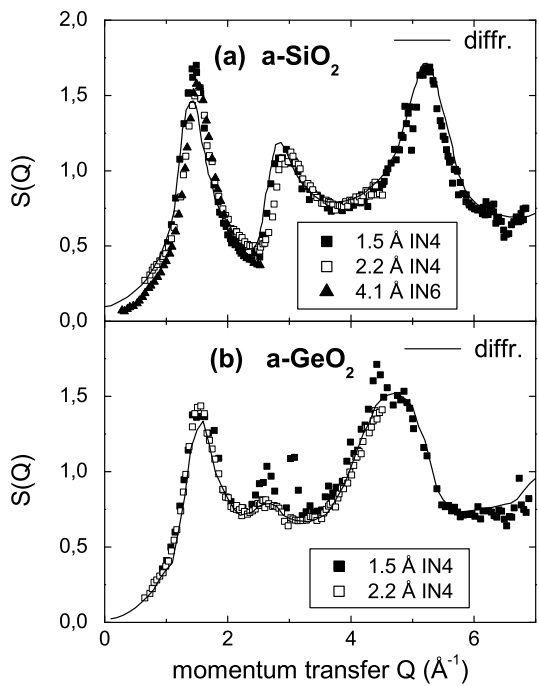,width=8
cm,angle=0} \vspace{0cm} \caption{(a) Normalization of IN4 data at
2.2 ${\rm\AA}$ and 1.5 ${\rm\AA}$ to the $S(Q)$ of vitreous silica
\cite{wright,stone} (continuous line). An earlier IN6 measurement
\cite{wischi} at 4.1 ${\rm\AA}$ is also included (b) Normalization
of IN4 data at 2.2 ${\rm\AA}$ and 1.5 ${\rm\AA}$ to the $S(Q)$ of
vitreous germania \cite{suzuya} (continuous line).}
\end{figure}
%%%%%%%%%%%%%%%%%%%%% end figure %%%%%%%%%%%%%%%%%%%%%%%%%%%%%%%%%%%%%%%

The procedure is not exact, because the coherent inelastic
scattering at small momentum transfer is not exactly zero. If the
velocity of the incoming neutrons exceeds the longitudinal sound
velocity of the glass (the kinematic condition \cite{lovesey} for
the visibility of the Brillouin line), one sees the Brillouin
scattering. But even if it does not (as in the measurements
reported here), there is still a small nonzero coherent inelastic
scattering contribution. This, however, is small compared to the
multiple scattering \cite{russina,schmidt}.

\subsection{Normalization to $S(Q)$}

In principle, if one does all corrections properly, one should be
able to normalize the measurement to the vanadium signal. In
practice, it is easier and more accurate to normalize the
time-of-flight data of a glass to the $S(Q)$ of a diffraction
measurement. For vitreous silica, there are two such diffraction
measurements in the literature \cite{wright,stone}. They agree
very well with each other. For germania, there is only one
diffraction measurement \cite{suzuya}.

%%%%%%%%%%%%%%%%%%%%% begin figure %%%%%%%%%%%%%%%%%%%%%%%%%%%%%%%%%%%%%
\begin{figure}[b]
\hspace{-0cm} \vspace{0cm} \epsfig{file=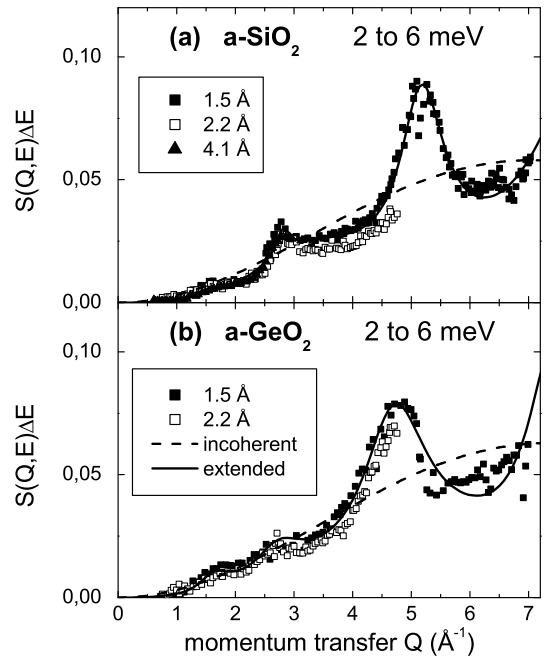,width=8
cm,angle=0} \vspace{0cm} \caption{The inelastic dynamic structure
factor in the boson peak region from (a) the three measurements of
vitreous silica (b) the two measurements of vitreous germania. The
dashed line is the incoherent approximation discussed in Section
III.A, the continuous line the extension to coherent scattering in
terms of the empirical five-lorentzian fit explained in Section
III.B.}
\end{figure}
%%%%%%%%%%%%%%%%%%%%% end figure %%%%%%%%%%%%%%%%%%%%%%%%%%%%%%%%%%%%%%%

To determine $S(Q)$ from a time-of-flight dataset, one integrates
the scattering over the time channels of a single detector or a
detector group, correcting for the change in momentum transfer as
the energy transfer changes. Here, we corrected with the factor
$Q_{el}^2/Q^2$, where $Q_{el}$ is the momentum transfer at the
elastic line and $Q$ is the one at finite energy transfer. The
procedure is not exact, but it has the advantage that a bad
detector does not corrupt its neighbors.

%%%%%%%%%%%%%%%%%%%%% begin figure %%%%%%%%%%%%%%%%%%%%%%%%%%%%%%%%%%%%%
\begin{figure}[b]
\hspace{-0cm} \vspace{0cm} \epsfig{file=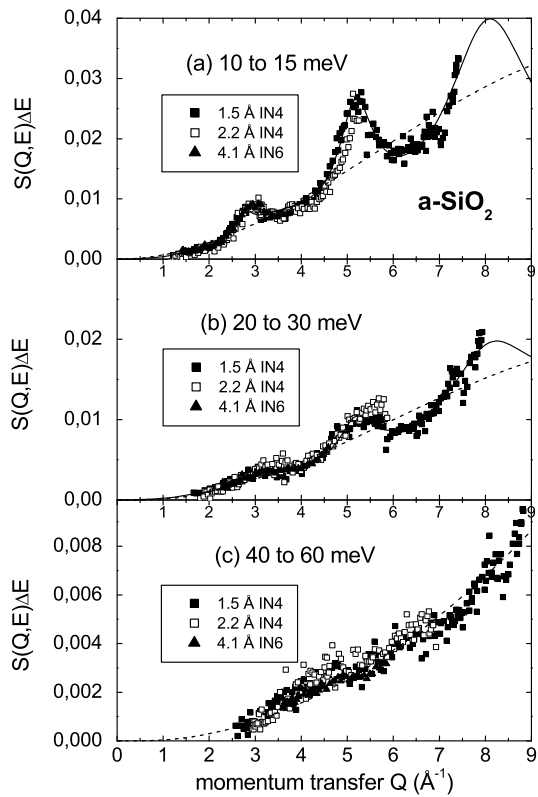,width=8
cm,angle=0} \vspace{0cm} \caption{The development of the inelastic
dynamic structure factor of vitreous silica above the boson peak.
The dashed line is the incoherent approximation discussed in
Section III.A, the continuous line the extension to coherent
scattering in terms of eq. (\ref{sd}). (a) Energy transfer from 10
to 15 meV (b) Energy transfer from 20 to 30 meV (c) Energy
transfer from 40 to 60 meV.}
\end{figure}
%%%%%%%%%%%%%%%%%%%%% end figure %%%%%%%%%%%%%%%%%%%%%%%%%%%%%%%%%%%%%%%

Fig. 3 (a) shows the result of this normalization to $S(Q)$ for
the two IN4 measurements on vitreous silica at room temperature.
Also included is an earlier IN6 measurement \cite{wischi}, done
with incoming neutrons of a wavelength 4.1 ${\rm\AA}$ at 318 K.

The same normalization of the two IN4 room temperature
measurements of vitreous germania to diffraction data from this
substance \cite{suzuya} is shown in Fig. 3 (b). The 1.5 ${\rm\AA}$
measurement suffers from problems with the subtraction of the
empty container; it consisted of pure aluminum with large
crystalline grains.

The comparison to Fig. 3 (a) shows that the second and the third
peak in $S(Q)$ shift to lower momentum transfer in germania. This
is caused by the increased Ge-O distance (1.73 $~\AA$) in germania
\cite{suzuya} as compared to the Si-O distance of 1.6 $~\AA$ in
silica \cite{wright}. The corner-connected GeO$_4$-tetrahedra in
germania are larger than the SiO$_4$ ones in silica. Nevertheless,
the position of the first sharp diffraction peak is more or less
the same in both substances, which shows that the packing of the
tetrahedra must be different.

The normalization allows to compare different measurements in the
same frequency window. Fig. 4 (a) shows the boson peak frequency
window from 2 to 6 meV in vitreous silica for all three sets of
data. One sees pronounced oscillations around the dashed line
(calculated from the incoherent approximation as explained in
Section III.A). Germania shows rather similar oscillations in the
same window in Fig. 4 (b) (its boson peak is also at $\approx$ 4
meV). The task of the next section is to extract the information
content in these oscillations.

As one goes up in energy transfer, the amplitude of the
oscillations decreases gradually. This development is shown for
vitreous silica in Fig. 5 (a-c). Above 40 meV, one finds the
incoherent approximation to be valid within the experimental
error. In vitreous germania, the incoherent approximation is
reached even faster, around 30 meV. In both cases, there is no
discernible peak shift; within experimental error, the peaks
merely fade away with increasing energy transfer.

\section{Evaluation}

\subsection{Incoherent approximation}

Having data corrected for multiple scattering and normalized to
$S(Q)$, one can proceed to apply the incoherent approximation
described in Section II.B. In this way, one determines a
generalized vibrational density of states $g(E)$ (we replace the
frequency $\omega$ by the energy transfer $E=\hbar\omega$ in this
section) without any adaptable parameter.

In both cases, silica and germania, the procedure described in
section II.B converged to a final generalized vibrational density
of states after three iteration steps.

%%%%%%%%%%%%%%%%%%%%% begin figure %%%%%%%%%%%%%%%%%%%%%%%%%%%%%%%%%%%%%
\begin{figure}[b]
\hspace{-0cm} \vspace{0cm} \epsfig{file=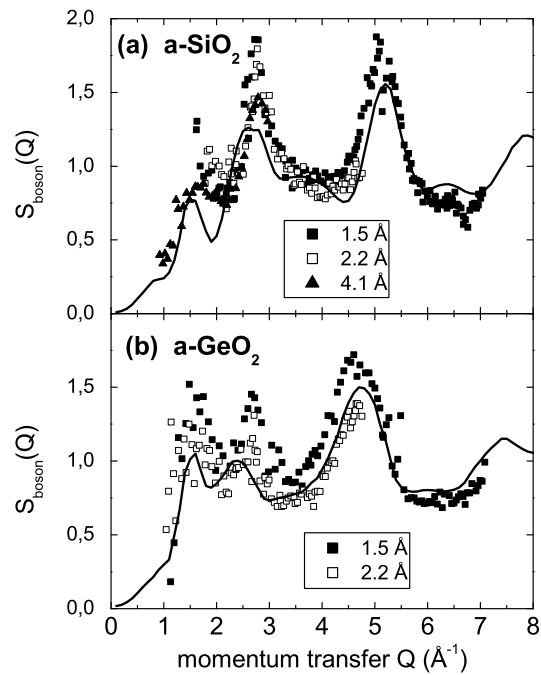,width=8
cm,angle=0} \vspace{0cm} \caption{Inelastic dynamic structure
factor $S_{boson}(Q)$ in the boson peak region for (a) silica
(continuous line sum of 0.45 $S(Q)$ and 0.55 tetrahedra libration,
see text) (b) germania (continuous line sum of 0.75 $S(Q)$ and
0.25 tetrahedra libration).}
\end{figure}
%%%%%%%%%%%%%%%%%%%%% end figure %%%%%%%%%%%%%%%%%%%%%%%%%%%%%%%%%%%%%%%

As it turns out, the three results for $g(E)$ of vitreous silica
agree fairly well with each other. The boson peak in $g(E)/E^2$ is
slightly lower in the 2.2 ${\rm\AA}$ measurement, but this is not
due to the oscillations of the inelastic dynamic structure factor.
Looking at Fig. 4 (a), one sees that the normalized intensity
itself is slightly lower than in the two other measurements.

The same good agreement was found for the vibrational densities of
states of germania calculated for the IN4 measurements at 1.5 and
2.2 ${\rm\AA}$.

\subsection{Analysis of the inelastic dynamic structure factor}

As the following analysis shows, we have at present no perfect
eigenmode fit of the dynamic structure factor at the boson peak.
Therefore we do the analysis in two steps. We first fit the
oscillation function $S_{boson}(Q)$ at the boson peak in a purely
empirical way with five lorentzians, independent of any motional
model, but providing an excellent fit. This fit form is used to
investigate the frequency dependence of the peaks in $S_\omega(Q)$
to higher frequencies. The second step models the atomic motion at
the boson peak in terms of a sum of translation and rotation of
undistorted tetrahedra. In this procedure, one has only one fit
parameter (the translational fraction of the motion). One gets a
reasonable rather than a perfect fit, but one has a motional model
which allows to calculate the true rather than the generalized
vibrational density of states.

To go beyond the incoherent approximation, we start by calculating
the oscillation function $S_\omega(Q)$ at the boson peak via
\begin{equation}\label{sboson}
S_{boson}(Q)= \frac{S_{boson}(Q,\omega)}{1/\pi\int_0^\infty
dt\cos\omega t{\rm e}^{-\gamma(t)Q^2}},
\end{equation}
taking $\gamma(t)$ from the evaluation in terms of the incoherent
approximation described in the previous subsection.
$S_{boson}(Q,\omega)$ is determined from the measured
$S(Q,\omega)$ via
\begin{equation}\label{sqomboson}
S_{boson}(Q,\omega)=\frac{\int_{\omega_{min}}^{\omega_{max}}S(Q,\omega)
d\omega}{\omega_{max}-\omega_{min}},
\end{equation}
where $\hbar\omega_{min}$=2 meV was chosen to exclude any
contribution of the elastic line even for the 1.53 ${\rm\AA}$
measurement and $\hbar\omega_{max}$=6 meV is a frequency well
above the two boson peak frequencies of 4 meV and 3.6 meV for
silica and germania, respectively. Fig. 6 (a) shows the result for
the three sets of data of silica, Fig. 6 (b) for the two IN4
measurements of germania.

%%%%%%%%%%%%%%%%%%%%% begin figure %%%%%%%%%%%%%%%%%%%%%%%%%%%%%%%%%%%%%
\begin{figure}[b]
\hspace{-0cm} \vspace{0cm} \epsfig{file=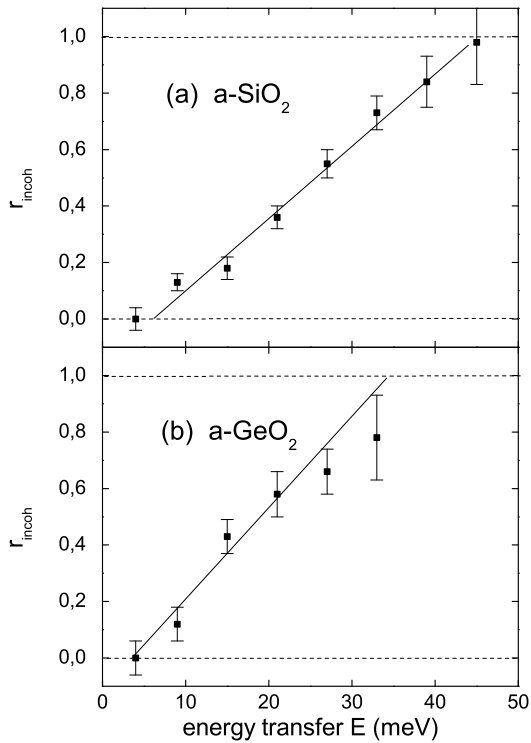,width=8
cm,angle=0} \vspace{0cm} \caption{The change of the incoherent
fraction $r_{incoh}$ of the oscillation function with increasing
frequency for (a) vitreous silica (b) vitreous germania.}
\end{figure}
%%%%%%%%%%%%%%%%%%%%% end figure %%%%%%%%%%%%%%%%%%%%%%%%%%%%%%%%%%%%%%%

The next step is to quantify the fading-away of the oscillations
with increasing frequency. For this, we need a functional
expression for $S_{boson}(Q)$. To get it, we fit the data points
of Fig. 6 (a) in terms of a sum of five lorentzians (the
continuous line in Fig. 6 (a)). This purely empirical function
oscillates around 1 in the momentum transfer region of the three
measurements. One can use it to replace the incoherent
approximation $S_{boson}(Q)=1$. Fig. 4 (a) shows that it gives a
much better agreement with experiment than the incoherent
approximation, at least in the frequency region of the boson peak.

%%%%%%%%%%%%%%%%%%%%% begin figure %%%%%%%%%%%%%%%%%%%%%%%%%%%%%%%%%%%%%
\begin{figure}[b]
\hspace{-0cm} \vspace{0cm} \epsfig{file=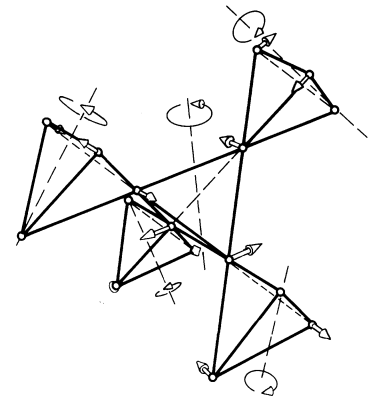,width=5
cm,angle=0} \vspace{0cm} \caption{Coupled libration of five
corner-connected SiO$_4$- or GeO$_4$-tetrahedra \cite{bu}.}
\end{figure}
%%%%%%%%%%%%%%%%%%%%% end figure %%%%%%%%%%%%%%%%%%%%%%%%%%%%%%%%%%%%%%%

As one goes up in frequency, the oscillations of the measured
dynamic structure factor begin to get weaker, until one reaches
again the incoherent approximation at about 40 meV (see Figs. 5(a)
to (c)). One can follow this behaviour quantitatively by fitting
the inelastic intensities in subsequent frequency windows in terms
of the oscillation function
\begin{equation}\label{sd}
S_\omega(Q)=r_{incoh}(E)+\left[1-r_{incoh}(E)\right]S_{boson}(Q)
\end{equation}
where $S_{boson}(Q)$ is the five-lorentzian fit of Fig. 4 and
$r_{incoh}(E)$ is the energy-dependent fraction of incoherent
scattering, going from 0 to 1 as one goes from the boson peak up
to higher frequencies (Fig. 7 (a)). Fig. 5 shows that one obtains
a good fit of the measurements in this way. So within experimental
accuracy, the peaks in the coherent inelastic dynamic structure
factor of silica do not shift; they merely fade away to make room
for a full validity of the incoherent approximation above 40 meV.

%%%%%%%%%%%%%%%%%%%%% begin figure %%%%%%%%%%%%%%%%%%%%%%%%%%%%%%%%%%%%%
\begin{figure}[b]
\hspace{-0cm} \vspace{0cm} \epsfig{file=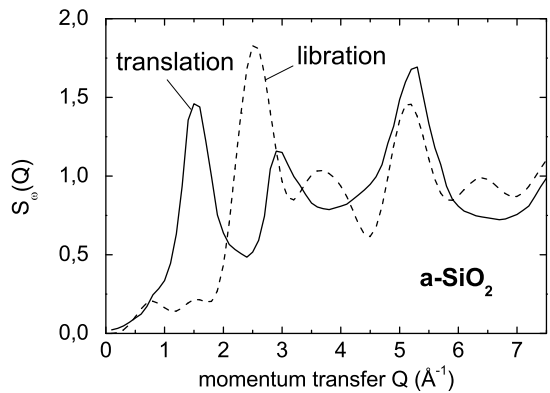,width=8
cm,angle=0} \vspace{0cm} \caption{Comparison of the translation
oscillation function $S(Q)$ (continuous line) with the oscillation
function $S_{rot}(Q)$ (dashed line) of the coupled libration of
five corner-connected SiO$_4$-tetrahedra \cite{bu} of Fig. 8.}
\end{figure}
%%%%%%%%%%%%%%%%%%%%% end figure %%%%%%%%%%%%%%%%%%%%%%%%%%%%%%%%%%%%%%%

It is interesting to note that this crossover into a validity of
the incoherent approximation occurs at the Debye frequency
$\omega_D$ of vitreous silica. Debye's strongly oversimplified
picture describes the whole vibrational density of states in terms
of sound waves. The total density of sound waves (normalized to 1)
is
\begin{equation}\label{debye}
g(\omega)=\frac{3\omega^2}{\omega_D^3},
\end{equation}
with the Debye frequency $\omega_D$ given by
\begin{equation}
\omega_D^3=\frac{18\pi^2\rho}{\overline{M}(1/v_l^3+2/v_t^3)}.
\end{equation}
Here $\rho$ is the density, $v_l$ is the longitudinal sound
velocity and $v_t$ is the transverse sound velocity. Table I gives
the values for vitreous silica and germania.

\begin{table}
\caption{Debye frequencies of vitreous silica and germania
\cite{zepo}.}
\bigskip
\begin{tabular}{|c|c|c|c|c|c|}
  \hline
  substance & $\overline{M}$ & $\rho$ & $v_l$ & $v_t$ & $\hbar\omega_D$ \\
   & a.u. & $kg/m^3$ & $m/s$ & $m/s$ & $meV$ \\
  $a-SiO_2$ & 20 & 2200 & 5800 & 3800 & 42 \\
  $a-GeO_2$ & 34.86 & 3600 & 3680 & 2410 & 26.8 \\ \hline
\end{tabular}
\end{table}

In germania, the oscillations of the inelastic dynamic structure
factor fade even more quickly with increasing frequency than in
silica (Fig. 7 (b)). Again, the validity of the incoherent
approximation is reached at about the Debye frequency.

Having established the frequency dependence of the dynamic
structure factor in both glasses, we proceed to the second step,
the understanding of the oscillation function $S_{boson}(Q)$ at
the boson peak in terms of a detailed picture of the atomic
motion.

To find such an understanding, let us first recall what one knows
about this frequency region, both from the quartz crystal
\cite{grimm,strauch} and from the silica glass neutron
\cite{bu,wischi} and Hyperraman \cite{hehlen} studies. One expects
a mixture of long-wavelength sound waves and SiO$_4$-tetrahedra
libration. The oscillation function of the translational motion of
long-wavelength sound waves \cite{carp} is $S(Q)$. The oscillation
function $S_{rot}(Q)$ of the coupled tetrahedra libration was
calculated from the motional model of Fig. 8, using eq.
(\ref{somq1}) for $S_\omega(Q)$ in the one-phonon approximation.

Fig. 9 compares the two oscillation functions $S(Q)$ and
$S_{rot}(Q)$ for silica. They differ mainly at the first and
second peak. Thus a strong first peak in $S_{boson}(Q)$ means a
large sound wave fraction in the boson peak modes, a strong second
peak a large tetrahedra libration fraction.

%%%%%%%%%%%%%%%%%%%%% begin figure %%%%%%%%%%%%%%%%%%%%%%%%%%%%%%%%%%%%%
\begin{figure}[b]
\hspace{-0cm} \vspace{0cm} \epsfig{file=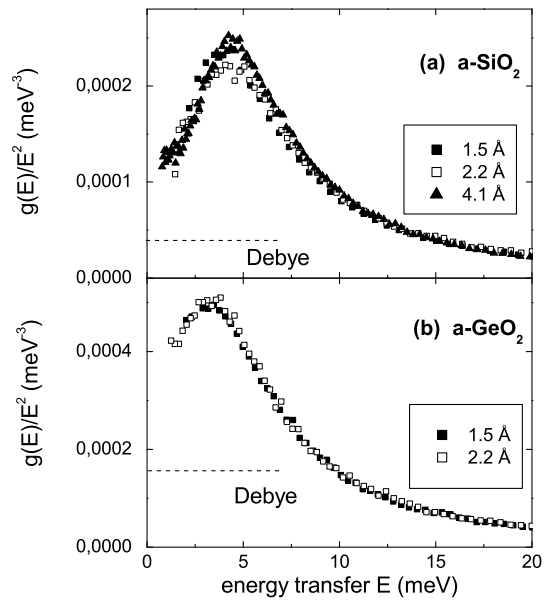,width=8
cm,angle=0} \vspace{0cm} \caption{Vibrational density of states
(plotted as $g(E)/E^2$ to emphasize the boson peak at 4 $meV$),
obtained using the extended approximation for (a) the three
time-of-flight datasets of vitreous silica (dashed line: Debye
expectation) (b) the two IN4 measurements of vitreous germania
(dashed line: Debye expectation).}
\end{figure}
%%%%%%%%%%%%%%%%%%%%% end figure %%%%%%%%%%%%%%%%%%%%%%%%%%%%%%%%%%%%%%%

Fig. 6 (a) shows the best fit (continuous line) of the observed
$S_{boson}(Q)$ of silica in terms of a sum of the two oscillation
functions. The fit is by no means perfect, but supplies a fraction
of 0.55 of tetrahedra libration signal and 0.45 of translation.

This is a surprising result. If one estimates the strength of the
translational signal on the basis of the Debye sound-wave
treatment, one would expect no more than a translational fraction
of 0.2 at the boson peak \cite{bu,wischi}. Similarly, the
Hyperraman data \cite{hehlen} require a dominating role of the
tetrahedra libration for their understanding. But obviously, the
librational motion of the corner-connected tetrahedra is
accompanied by strong translational shifts.

A similar effect appears in germania. The fit in terms of a sum of
the $S(Q)$ of germania and the tetrahedra libration (the line in
Fig. 6 (b)) gives an even larger fraction of 0.75 for $S(Q)$. As
we will see, the Debye expectation in germania is again a factor
of two smaller. We will come back to this point in the discussion.

\subsection{Vibrational density of states in the extended approximation}

If one knows what kind of vibrational modes one deals with, one
can calculate the vibrational density of states in the extended
approximation, accounting for the cross sections and amplitudes of
the atoms participating in the modes.

For vitreous silica, we take the signal at the boson peak to
consist of 55 \% of tetrahedra libration signal and 45 \% of
$S(Q)$, the result of the best fit of the data in Fig. 6 (a).
Since the tetrahedra libration is essentially oxygen motion, we
have to correct the number of resulting modes dividing by the
enhancement factor for a pure oxygen motion
\begin{equation}\label{enhance}
f_O=\frac{\sigma_O}{\overline{\sigma}}\frac{\overline{M}}{M_O}=1.49,
\end{equation}
where $\sigma_O$ is the oxygen cross section and $M_O$ is the
oxygen mass. $\overline{\sigma}$ and $\overline{M}$ are the
average values of cross section and mass as defined in Section II.

The increasing incoherent fraction $r_{incoh}$ of $S_{dyn}(Q)$ in
eq. (\ref{sd}) is taken to be an average motion of all atoms,
which requires no enhancement factor.

With these assumptions, one can determine a vibrational density of
states from the data using the extended approximation, eq.
(\ref{extend}). $S_{boson}(Q)$ is calculated from the sum of 45 \%
of $S(Q)$ and 55 \% of the tetrahedra libration oscillation
function $S_{rot}(Q)$ of the model in Fig. 8. $S_\omega(Q)$ is
evaluated from eq. (\ref{sd}), taking $r_{incoh}$ to follow the
line in Fig. 7.

Fig. 10 (a) shows the vibrational density of states of vitreous
silica  obtained in this way for the three sets of data, plotted
as $g(E)/E^2$ to emphasize the boson peak at 4 $meV$.

For vitreous germania, we take the signal at the boson peak to
consist of 25 \% of tetrahedra libration signal and 75 \% of
$S(Q)$. Again, we have to correct the number of resulting modes
dividing by the enhancement factor for a pure oxygen motion
\begin{equation}\label{enhance2}
f_O=\frac{\sigma_O}{\overline{\sigma}}\frac{\overline{M}}{M_O}=1.59.
\end{equation}
The resulting vibrational density of states $g(E)/E^2$ is shown in
Fig. 10 (b).

\section{Discussion}

\subsection{Potential and limitations of the method}

Let us begin the general discussion with a disclaimer: The
evaluation method for coherent inelastic neutron or x-ray
scattering from glasses proposed in the present paper, which we
denote by "extended approximation", is by no means new. Its
beginnings can be traced back to the classical paper of Carpenter
and Pelizzarri \cite{carp} (and even beyond that). In a less
formal way, it has been applied earlier, not only to vitreous
silica \cite{bu,wischi}, but also to amorphous germanium
\cite{prager}, deuterated polybutadiene \cite{pbd} and boron
trioxyde \cite{engberg}. Our present work merely formalizes this
extended approximation, introducing the concept of the oscillation
function $S_\omega(Q)$ and giving a recipe for its experimental
determination.

The oscillation function $S_\omega(Q)$ contains information on the
eigenvector of the vibrational or relaxational modes seen at the
frequency $\omega$. The information is limited: Even if one knows
$S_\omega(Q)$ with high accuracy over a large $Q$-range, one
cannot determine the eigenvectors at that frequency. One can only
check models of the motion against the measured $S_\omega(Q)$. In
a practical sense, even such a check is restricted to the use of
the one-phonon approximation of eq. (\ref{somq1}), because it is
difficult to calculate the interference oscillations of the
coherent multiphonon scattering. Fortunately, the one-phonon
approximation holds to rather large momentum transfer at the boson
peak, the most interesting target for these studies.

At the boson peak, the analysis is simplified by the scientific
question: one wants to know (i) whether the long-wavelength sound
waves at the boson peak frequency still follow the Debye
expectation (ii) whether there are additional modes, and if there
are, what their eigenvector is. The first of these questions can
be attacked, using the fact \cite{carp} that the oscillation
function $S_\omega(Q)\Rightarrow S(Q)$ for long-wavelength sound
waves, no matter whether they are transverse or longitudinal. The
second requires a model calculation for whatever mode is expected
to be soft in the given system. In our cases silica and germania,
one expects the coupled libration of corner-connected tetrahedra
to be soft, because this is the soft mode of the phase
transformation from $\alpha$ to $\beta$ in crystalline quartz
\cite{grimm}.

%%%%%%%%%%%%%%%%%%%%% begin figure %%%%%%%%%%%%%%%%%%%%%%%%%%%%%%%%%%%%%
\begin{figure}[b]
\hspace{-0cm} \vspace{0cm} \epsfig{file=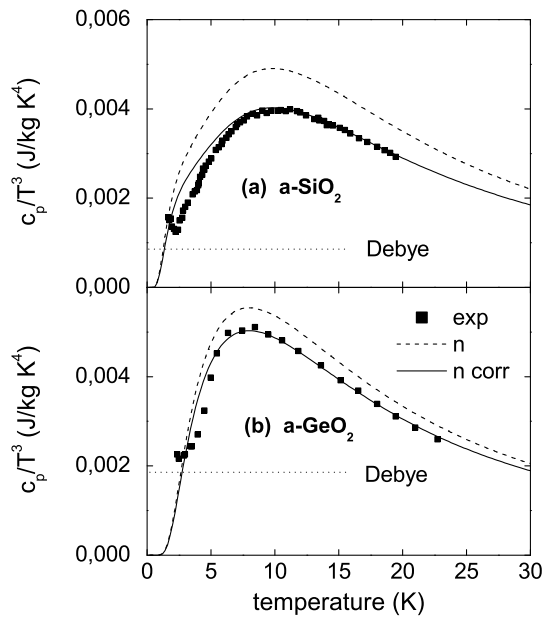,width=8
cm,angle=0} \vspace{0cm} \caption{Comparison of the heat capacity
calculated from the neutron data with and without correction for
the enhancement factor of the tetrahedra librational modes (the
one without correction is close to the result from the incoherent
approximation) to measured data (a) of vitreous silica \cite{bu}
(b) of vitreous germania \cite{antoniou}.}
\end{figure}
%%%%%%%%%%%%%%%%%%%%% end figure %%%%%%%%%%%%%%%%%%%%%%%%%%%%%%%%%%%%%%%

Note that this rather streamlined procedure has weak points: (1)
since the sound waves interact with the additional modes
\cite{schober1,schober2}, one needs the $S_\omega(Q)$ of the
resulting true eigenmodes, in which translation and soft mode
eigenvector parts have a nonzero interference term. Therefore
their $S_\omega(Q)$ might be different from a simple sum of the
two oscillation functions. (2) At the boson peak, measurements of
incoherent scatterers \cite{kanaya,pecha} reveal sizeable
nongaussianity effects. These will tend to distort the
experimental $S_\omega(Q)$, determined on the basis of the
Gaussian approximation.

In view of these points, it is not surprising that the agreement
between the model calculation and the data in Fig. 6 (a) and (b)
is not perfect. Nevertheless, the comparison with heat capacity
data in the next subsection shows the reliability of the resulting
vibrational density of states.

The extended approximation should be particularly useful for the
evaluation of wide-angle inelastic x-ray scattering data. Such
measurements have been reported \cite{masciovecchio,pilla}, but
have not been evaluated on a quantitative level. With the recipe
given here, one could do a quantitative evaluation, provided one
has good x-ray diffraction measurements. In that case, one even
could do only constant-Q scans (simpler to measure than the
constant-E scans of the two references
\cite{masciovecchio,pilla}), relating their intensities by the
diffraction measurement.

\subsection{Comparison to heat capacity}

We want to check the vibrational density of states obtained in the
previous section from the extended approximation against earlier
results in the literature. The first check is against heat
capacity data from silica \cite{bu} and germania \cite{antoniou}
between 1 and 20 K. If one plots the heat capacity $C_p$ as
$C_p/T^3$, one gets a close correspondence to the plot of
$g(E)/E^2$, showing the boson peak of silica and germania at about
10 K.

%%%%%%%%%%%%%%%%%%%%% begin figure %%%%%%%%%%%%%%%%%%%%%%%%%%%%%%%%%%%%%
\begin{figure}[b]
\hspace{-0cm} \vspace{0cm} \epsfig{file=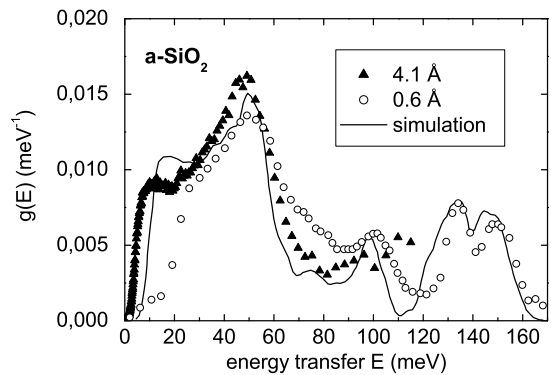,width=8
cm,angle=0} \vspace{0cm} \caption{Vibrational density of states of
vitreous silica from an {\it ab initio} simulation \cite{kob},
from earlier neutron work \cite{price} with 0.6 ${\rm\AA}$ and
from the present evaluation of the 4.1 ${\rm\AA}$ data.}
\end{figure}
%%%%%%%%%%%%%%%%%%%%% end figure %%%%%%%%%%%%%%%%%%%%%%%%%%%%%%%%%%%%%%%

In this comparison, the mode eigenvector assignment at the boson
peak is checked, because the assumed fraction of tetrahedra
libration provides the correction to the number of vibrational
modes.

Fig. 11 (a) compares measured heat capacity data \cite{bu} of
vitreous silica to the result of the evaluation of the neutron
data in the extended approximation, described in the preceding
section. We chose the results from the measurement at 4.1
${\rm\AA}$, because the heat capacity measurements \cite{bu} stem
from the same sample. As it turns out, the correction of the
enhancement factor of eq. (\ref{enhance}) is essential to get good
agreement. So the comparison corroborates the assignment of 55 \%
pure oxygen signal and 45 \% translational signal at the boson
peak in vitreous silica.

Similarly, Fig. 11 (b) shows good agreement between measured
\cite{antoniou} and calculated heat capacity data in vitreous
germania. Here, we compare to the 2.2 ${\rm\AA}$ measurement,
because it has the better resolution. Again, the correction
improves the agreement. However, in this second case the
correction is smaller, because according to the fit of the
measured dynamic structure factor of Fig. 6 (b) we have only 25 \%
tetrahedra libration at the boson peak. Again, this conclusion is
supported by the heat capacity data.

\subsection{Comparison to simulation}

There is a large number of molecular-dynamics simulations of
vitreous silica in the literature\cite{taraskin,delanna,horbach},
most of them done with effective classical force fields like the
BKS potential \cite{beest}. The BKS potential reproduces the
measured $S(Q)$, but a recent comparison to an {\it ab initio}
calculation \cite{kob} shows that it fails not only to reproduce
the vibrational density of states, but also the mode eigenvectors.

Fig. 12 compares the vibrational density of states determined from
the neutron measurement at 4.1 ${\rm\AA}$ in the extended
approximation with the {\it ab initio} calculation \cite{kob} and
with an earlier neutron determination with very short wavelength
at the spallation source at Argonne \cite{price}. Note that the
two neutron determinations are complementary: the spallation
source measurement suffers from an overcorrection at low
frequencies, but provides a true picture of the high-frequency
density of states. In contrast, the cold neutron measurement is
unable to measure above 100 meV, but provides good results at low
frequency, as shown by the comparison to the low-temperature heat
capacity data in the previous subsection. Between 20 and 100 meV,
the two sets of data agree reasonably well with each other.

Note that the {\it ab initio} simulation is unable to describe the
boson peak region below 10 meV. Above that region, there is
impressive agreement between simulation and neutron experiment.
But below 10 meV, the simulated modes fail to come down to the low
frequencies where both neutrons and heat capacity report them to
exist. The reason might be either the small size of the simulation
cell or its poor equilibration, natural disadvantages of an {\it
ab initio} simulation.

It is interesting to compare our results to a simulation
\cite{schober2} of dynamic structure factors of a different
system, soft spheres interacting with a repulsive 1/r$^6$
potential, a model appropriate for metallic glasses. In this case,
the oscillation function at the boson peak shows only the peaks of
$S(Q)$. The additional modes seem to be a motion along a chain of
nearest atomic neighbors, with an oscillation function which
resembles $S(Q)$. This is obviously different in silica and
germania, where the resonant boson peak modes seem to have a
distinct tetrahedra rotation component.

\subsection{Sound waves at and above the boson peak}

The preceding two subsections demonstrated the ability of the
extended approximation to determine a reliable vibrational density
of states, in particular in the frequency region of the boson
peak. This good agreement supports the interpretation of the boson
peak modes as a mixture between long-wavelength sound waves and
soft modes, in the cases of silica and germania librational modes
of corner-connected tetrahedra.

%%%%%%%%%%%%%%%%%%%%% begin figure %%%%%%%%%%%%%%%%%%%%%%%%%%%%%%%%%%%%%
\begin{figure}[b]
\hspace{-0cm} \vspace{0cm} \epsfig{file=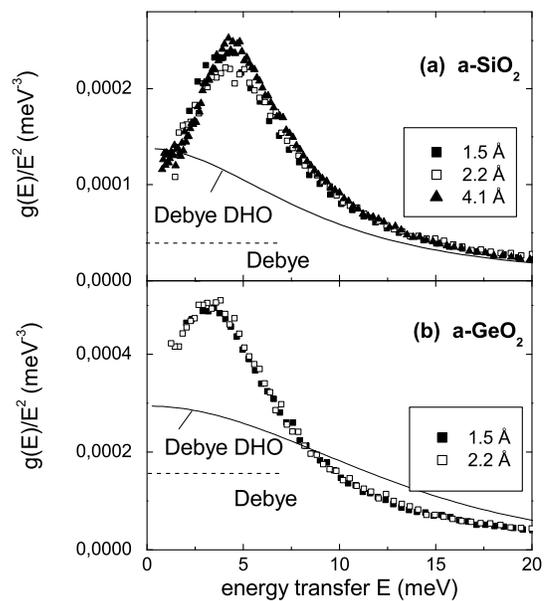,width=8
cm,angle=0} \vspace{0cm} \caption{Experimental density of states,
plotted as $g(E)/E^2$ as in Fig. 10, compared to the simple Debye
expectation (dashed line) and to Debye sound waves with DHO
damping as explained in the text (continuous line) for (a)
vitreous silica (b) vitreous germania.}
\end{figure}
%%%%%%%%%%%%%%%%%%%%% end figure %%%%%%%%%%%%%%%%%%%%%%%%%%%%%%%%%%%%%%%

But the sound-wave fraction determined from the dynamic structure
factor is decidedly higher than the expectation on the basis of
the sound-wave Debye model. Looking at the boson peak region
between 2 and 6 meV in Fig. 10 (a) and (b) and the dashed line of
the Debye expectation, one would expect no more than 20 \% sound
waves in silica and no more than 40 \% in germania. The fit of
$S_{boson}(Q)$ gives about twice as much in both substances. This
finding goes beyond earlier cold-neutron work \cite{bu,wischi},
which could not quantify the sound wave fraction at the boson
peak.

The effect is apparently not limited to silica and germania.
Earlier decompositions of $S(Q,\omega)$ at the boson peak window
into a $Q^2$- and a $Q^2S(Q)$-part in deuterated polybutadiene
\cite{pbd} and in vitreous $B_2O_3$ \cite{engberg} also observed a
larger $Q^2S(Q)$-part than expected on the basis of the Debye
model.

If the $Q^2S(Q)$ part is only due to sound waves, this implies a
downward shift in frequency of the sound wave intensity above the
boson peak. For the longitudinal sound waves, such a downward
shift, together with a strong broadening, is in fact observed
experimentally in x-ray Brillouin scattering \cite{benassi}. The
broadening and the shift increase with the square of the phonon
wavevector, i.e. with the square of the nominal frequency of the
phonon.

There is general agreement that the downward shift is not due to a
dispersion of the sound velocity, but rather to sound wave
scattering. One finds no evidence for any dispersion of either the
longitudinal or the transverse sound velocity in tunneling diode
\cite{rothenfusser} or ballistic phonon-pulse \cite{maris}
experiments in vitreous silica and other glasses.

Also, there seems to be general agreement that the broadening of
the sound waves observed in x-ray Brillouin scattering is not due
to a true anharmonic damping of the sound waves, but rather to a
deviation of the true eigenvectors from a purely sinusoidal
displacement in space \cite{delanna}.

On the other hand, there is a hot debate on the proper description
of the x-ray Brillouin data, whether one should use a damped
harmonic oscillator \cite{benassi} (DHO) or whether one should
take another form \cite{rat} which brings no intensity down to the
frequency zero. But both forms shift the intensity down to lower
frequencies, consistent with our observation of a heightened
$Q^2S(Q)$-component at the boson peak (heightened with respect to
undamped Debye phonons).

The low-temperature plateau in the thermal conductivity
\cite{zepo} and the phonon-pulse experiments \cite{maris} suggest
that the transverse phonons have essentially the same fate as the
one observed for the longitudinal phonons in x-ray Brillouin
scattering \cite{benassi,rat}.

To test these ideas, we submitted a Debye density of states with a
Debye frequency corresponding to an energy transfer of 42 meV (the
value for silica) to a DHO damping with the parameters of Benassi
{\it et al} \cite{benassi}. The transverse phonons were assumed to
have the same damping as the longitudinal ones at the same
frequency, increasing proportional to the square of the phonon
wavevector. The resulting effective Debye density of states is
shown as the continuous line in Fig. 13 (a). To do the same
calculation for germania, shown in Fig. 13 (b), we used recent
neutron Brillouin measurements \cite{geobrill}. According to them,
the strong damping condition $\Gamma=v_lq$ ($q$ phonon wavevector)
is reached at the energy transfer 19.3 meV, a bit less than three
quarters of the Debye frequency. From Benassi {\it et al}
\cite{benassi}, the same condition is reached in vitreous silica
already at 12 meV, at about only one quarter of the Debye
frequency.

Fig. 13 shows that the scattering of the sound waves is strong
enough to have a non-negligible effect on their spectral
appearance in a scattering experiment. But it also shows that one
cannot explain the boson peak in terms of the sound wave
scattering alone.  This does not depend on the choice of the DHO.
If one takes the alternative proposed by the Montpellier group
\cite{rat}, one does indeed get a peak, because this alternative
shifts the intensity to a finite frequency, not to the frequency
zero. But the intensity of the peak remains too small. What one
really needs is a mechanism which raises the total vibrational
mean-square displacement by nearly a factor of two over the Debye
expectation, a mechanism which brings vibrations from higher
frequencies down to the boson peak. In vitreous silica and
vitreous germania, one has the additional advantage that one can
identify these additional vibrations by their dynamic structure
factor.

\section{Summary}

We introduce a formal treatment of the coherent inelastic neutron
or x-ray wide-angle scattering from glasses, which takes the
interference oscillations explicitly into account. This "extended
approximation" is an extension of the incoherent approximation. It
allows to fit a newly defined "oscillation function" $S_\omega(Q)$
at each frequency $\omega$, thus supplying information on the
vibrational eigenvectors. The method should be particularly useful
for the quantitative evaluation of wide-angle inelastic x-ray
scattering measurements.

The application of the method to new room-temperature thermal
neutron time-of-flight measurements of silica and germania not
only provides a vibrational density of states in excellent
agreement with heat capacity and simulation data, but also allows
to quantify for the first time the ratio of tetrahedra rotation
and tetrahedra translation at the boson peak. One finds about
twice as much translation as in a simple Debye expectation. This
excess is probably connected to the heavy damping of the sound
waves at frequencies above the boson peak, observed by x-ray
Brillouin scattering.

\bigskip

\acknowledgements{Thanks are due to Herbert Schober and Andreas
Wischnewski for a critical reading of the manuscript.}

\end{document}